# The Development of Investment Planning Models for the United Kingdom's Wind and Solar Fleets

Anthony D Stephens and David R Walwyn
Correspondence to tonystephensgigg@gmail.com

**Abstract**

Previous work has resulted in the development of an energy model able to calculate wind and solar fleet efficiencies. However, for investment planning purposes, it is necessary to calculate from the lowest economically acceptable efficiencies how much wind and solar generation would be economically justified. The paper explains how this objective has been achieved with arrays (investment planning tables) created after carrying out a structured investigation of the behaviour of the electricity system over the whole of its operational range. The tables are then applied to National Grid prediction of the size and composition of the system in the year 2035. A conclusion is reached that wind and solar generation will only be able to supply about 70% of electrical demand, the other 30% being provided by dispatchable sources of generation, which must be sufficiently fast acting to maintain electricity system stability, such as the use of combined cycle gas turbines. This limit on deployment of wind and solar generation restricts their ability to decarbonise the electricity system and is likely to lead in 2035 to a residual of 71.7 Mtes per annum of carbon dioxide emissions which wind and solar generations will be unable to address.

**Keywords**

**Nomenclature**

| | |
|---|---|
| **GW** | Measure of generation in GW |
| **GW w** | Wind generation in GW |
| **GW s** | Solar generation in GW |
| **GW w+s** | Wind plus solar generation in GW accommodated by the electricity system |
| **Hdrm** | That portion of electrical demand available for wind/ solar generation to satisfy |
| **wm** | Multiple of wind generation of 6.045 GW in 2017 |
| **sm** | Multiple of solar generation of 1.16 GW in 2017 |
| **ccgt** | Combined cycle gas turbines |
| **IWE** | Incremental Wind Efficiency |
| **ISE** | Incremental solar Efficiency |

## 1. Introduction

Governments and their energy advisors must deal with an awkward but unavoidable difficulty in developing energy systems. Generation and transmission capacity takes years to design, install and commission, and yet future demand is uncertain and difficult to predict, being dependent on factors such as the uptake of battery electric vehicles, heat pumps and the replacement of natural gas by electricity in the manufacturing sector.

Models of the energy system are an important means by which the supply/demand balance can be assessed. This paper describes briefly one such model which operates at a low level of complexity, but nevertheless provides useful insights. These insights are illustrated with two important and relevant examples, firstly the appropriate level of wind fleet capacity in 2035 and secondly the complex relationship between wind and solar generation as both technologies become more significant to the overall generation mix.

The authors have previously published the development of a Compound Model whose purpose was to investigate the dynamic nature and efficiencies of wind and solar fleets powering future United Kingdom electricity systems (Stephens and Walwyn, 2023, 2020a, b). The Compound Model is based on appropriately scaled real time wind and solar generation records which are available on the internet, and its inclusion of 52 weekly dynamic models provides insights into the likely future dynamic nature of the electricity system for the whole range of meteorological conditions the wind and solar fleets are likely to encounter. As the wind and solar fleets increase in size, they inevitably start to exceed the needs of the electricity system and the weekly dynamic models reveal that excess generation is far too intermittent and variable in magnitude to be put to any beneficial use.

The curtailment of excess generation leads to a progressive reduction in the efficiency of the wind and solar fleets and eventually limits their economic deployment. Although the Compound Model can calculate these efficiencies, what is needed for investment planning purposes is to be able to calculate from the lowest economically acceptable efficiencies how much wind and solar generation would be economically justified. The paper explains how this objective has been achieved with arrays created after carrying out a structured investigation of the behaviour of the electricity system over the whole of its operational range. The result of the analysis can be used to calculate the economically justifiable size of the wind and solar generation in proportion to the total demand.

## 2. Background to Electricity Supply in the United Kingdom

The UK electricity supply industry experienced many fundamental changes in the way it operated in the second half of the 20[th] century (Burdon, 2010). The winter of 1947, the worst of the century, was the main stimulus for the nationalisation of the many small privately owned electricity generating companies. Nationalisation enabled the industry to establish a significant technical and scientific capability, both in personnel and R&D facilities, allowing prototypes to be built and assessed. Many were successful, leading to full scale deployment, while others such as the plutonium based Fast Breeder Reactor (FBR) proved the concepts to be impracticable. Between nationalisation in 1947 and privatisation in 1990 generation capacities increased some twentyfold and their efficiencies improved enormously as temperatures and pressures were increased.

The building of larger new power stations in the coal fields in the 1970s, together with the associated 400 KV transmission system, enabled "coal by wire" to be sent to the cities, which no longer had to suffer life - threatening smogs during the winter months. Following the successful commissioning in 1957 of Calder Hall, the world's first commercial nuclear power station, three generations of nuclear reactors were built; Magnox reactors, powered by unenriched uranium, and Advanced Gas Cooled

Reactors (AGRs) and Pressurised Water Reactors (PWRs), powered by low enriched uranium. At their peak, the nuclear reactors generated about a quarter of UK electrical demand.

The decision to build a fleet of AGRs by different consortia to different designs, turned out to be a disastrous one, compounded by their graphite moderators deteriorating over time and limiting their lives to 25-30 years. PWR's have typical lives of 60 years. By the end of the decade the UK is likely to have only the 1.2 GW Sizewell B PWR in operation, a situation which will significantly reduce the country's energy security.

The discovery of North Sea gas led to coal fired power stations being progressively displaced by more compact combined cycle gas turbines (ccgt) stations, which emit approximately half the carbon dioxide emitted by coal fired stations. Ccgts are ideally suited to load following, since they are able to increase their power output quickly. The remaining nuclear reactors have been given the highest priority access to the grid, since they have high capital costs but low operating cost, followed by wind and solar generation, with ccgts providing back up to ensure security of supply. All of the sources of generation employed in the 20$^{th}$ century, whether coal, nuclear or gas, are classified as "dispatchable", sources which may be directly controlled by the operator.

This has changed in the 21$^{st}$ century, with the introduction of wind and solar power which are highly dependent on meteorological conditions, and are not under the control of the operator. Until 2017 all wind and solar generation was accommodated by the electricity system, and it was possible to predict approximately their annual contributions from their average load factors; roughly 30% of capacity from the on- shore fleet, 40-45% of capacity from the off- shore fleet and 10% of capacity from the solar fleet. In April 2017 however, when wind fleet capacity was around 20 GW, the National Grid advised that the increasing size of the wind and solar fleets was leading to occasions when not all their generations could be accommodated by the electricity system. It had become necessary to pay some wind farms to cease generation (Gosden, 2017), and curtailment has increased substantially since 2017 (Gosden, 2021). Wind curtailment affects the efficiency of the wind fleet and the cost of generation, but its consequences are difficult to predict, since it takes place intermittently throughout the year.

This characteristic was the motivation for the development of a Compound Model, based on historic grid records, to predicting the efficiencies of future larger wind and solar fleets, and hence the upper economic limit of their deployment. It is important to differentiate between curtailment which arises because all available wind and solar generation cannot be accommodated by the electricity system, and that which arises because there is insufficient transmission capacity. A current government target is for there to be 50 GW of offshore capacity by 2030 (HM Government, 2023) but, if achieved, it appears unlikely that it will be possible to build by 2030 the additional transmission capacity which is estimated to cost £54Bn (Gosden, 2022). Curtailment due to lack of transmission capacity is not predictable and only curtailment caused by excess generation is included in our models.

3. **The Compound Model**

In this section, we cover the Compound Model and its use in calculating how much wind and solar generation the electricity system is able to accommodate, GWw+s, and two important wind and solar efficiencies, IWE and ISE

What made it possible to model the behaviour of future wind and solar fleets was the authors' previous finding that wind generation histograms for the years 2013-2016 were remarkably similar (Stephens and Walwyn, 2018). This explained the finding that when different years' records were appropriately scaled their predictions were also remarkably similar. Since solar generation must also be modelled, and solar generation records only became available on the internet in 2017 (Gridwatch, 2021), it was decided to base the Compound Model on 2017 records. A problem in handling a year's records is that the UK electricity system is recorded every five minutes, requiring 104,832 wind and solar records to be downloaded from the internet into the model. This problem was overcome by downloading the records a week at a time into 52 identical weekly sub-models, which are given identical input data by a coordinating spreadsheet, which also amalgamates their predictions to produce annual averages. An advantage of this approach is that the individual weekly model predictions may be graphed, providing valuable information about the likely dynamic behaviour of the future larger wind and solar fleets for the range of different meteorological conditions likely to be experienced.

A brief summary of what was concluded by studying the dynamic behaviour of three weekly models with different meteorological conditions appears in Appendix 1. Two important conclusions drawn were that generation in excess of the needs of the electricity system, excess generation, is far too intermittent and variable in magnitude to be put to any beneficial use, and that it will be necessary to retain a large fleet of ccgts to mitigate both extended wind lulls and rapid reductions in wind and solar generation. As has been increasingly the case since 2017, excess generation will be curtailed, and curtailment is calculated at each 5 minute interval in each of the 52 weekly sub- models by testing how much of the available wind plus solar generation the electricity system is able to accommodate. The weekly models calculate the weekly averages of accommodated wind plus solar generations, and these are averaged by the coordinating spreadsheet. The annual average wind plus solar generation accommodated by the electricity system is an important model parameter which we designate GW w+s.

What is needed when making investment decisions is how much of incremental increases in wind and solar generation are accommodated by the electricity system. We have called these the Incremental Wind Efficiency (IWE) and Incremental Solar Efficiency (ISE), and the Compound Model may be used to calculate IWE and ISE values in a two- stage process. The Compound Model is first used to make a base estimate of GW w+s, followed by two re-estimations of GW w+s for separate incremental increases of wind and solar generation. The differences between the base estimate and incrementally increased estimates, together with the sizes of the increments applied, enable the IWE and ISE values to be calculated. IWE and ISE are important model parameters because the unit costs of wind and solar generations are inversely proportional to their values, and the upper economic sizes of the wind and solar fleets will be mainly determined by the minimum economically acceptable IWE and ISE values.

## 4. Investment Tables

### 4.1 The Investment Planning Table

In 2017 the UK wind and solar fleets generated an average of 6.045 GW and 1.16 GW respectively. It is convenient to use multiples of these values as Compound Model input variables, and to designate them wm and sm. Nuclear generation has the highest access priority to the electricity system, followed by wind and solar generation. Electrical demand less nuclear generation is available to be satisfied by wind and solar generations, and we have called this Headroom, or Hdrm. The operating region of the wind and solar fleets may therefore be defined in terms Hdrm, wm and sm, which are the Compact Model input variables. One modelling priority is to assess the electricity system configuration in 2035, when the UK has a Climate Change target for greenhouse gas emissions to be 78% below their level in 1990 (Committee on Climate Change, 2020). In 2022 the National Grid published a scenario for the energy system in 2035 in FES 2022 (National Grid Electricity System Operator, 2022).

**Figure 1. Energy flow diagram in TWh proposed by the National Grid for 2035 in FES 2022[3]**

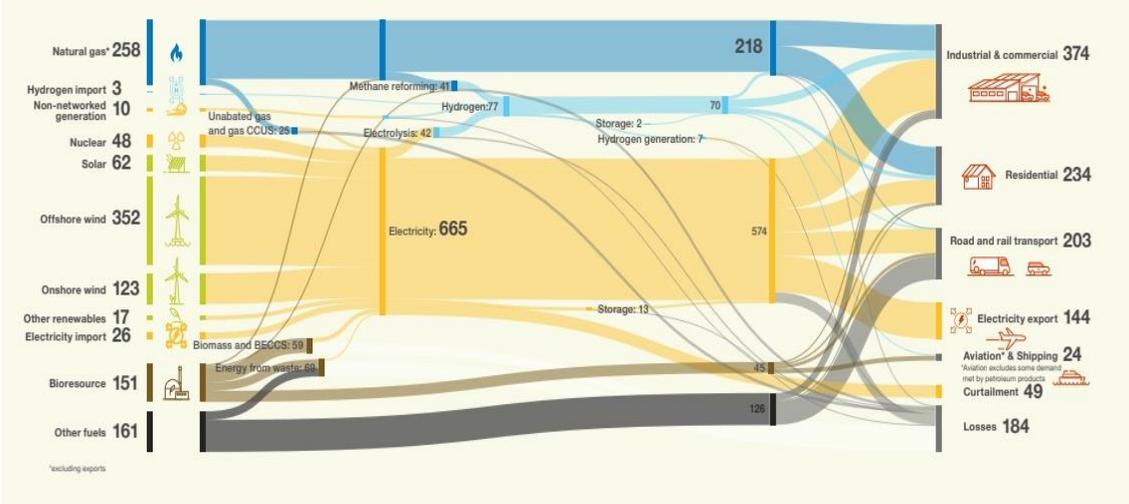

Figure 1 is a reproduction of the energy flow diagram for 2035 in FES 2022, in terms of annual TWh. Our interest is mainly in the electricity (yellow) component of Figure 1 which, with additional detailed information in a private communication with the National Grid, enabled electrical demands and sources of energy to be made in terms of annual average GW, leading to the Compound Model inputs for the 2035 scenario of Hdrm= 48.5, wm= 8.96 and sm= 6.1. Although, as described in the previous section, the Compound Model is able to calculate IWE and ISE values, for investment planning purposes it is the reverse of such calculation which are necessary; making calculations once minimum IWE and ISE values have been established of the wm and sm values which will satisfy these IWE and ISE values.

The approach adopted was to use the Compound Model to create an array of IWE values over the whole range of Hdrm, wm and sm values of interest which, with interpolation, may then be used to calculate the desired wm and sm values. In 2024 Hdrm is approximately 30 GW and, bearing in mind the anticipated Hdrm, wm and sm values for the 2035 scenario, the IWE array needed to explore the performance of the electricity system until 2035 must cover the electricity system operational range defined by Hdrm values of 30 to 50, wm values of 1 to 10 and sm values of 0 to 8. Adjacent entries in the IWE array are calculated at equal intervals of Hdrm, wm and sm, and care must be taken when choosing the intervals.

The IWE vs wm relationship is significantly non- linear in some regions of the operational area of interest, leading to the danger of computational errors if the intervals are too large. On the other hand, the number of IWE calculations needed is the product of the numbers of Hdrm, wm and sm values assessed, with the danger of unnecessary computational complexity if the intervals chosen are too small. In order to assess the appropriately sized intervals the Compound Model was used to generate the IWE vs wm curves of Figure 2.

**Figure 2. IWE vs wm for Hdrm=50, sm=0 to 8 (left), and and Hdrm=30, 40, 50 &60 (for sm=6) (right)**

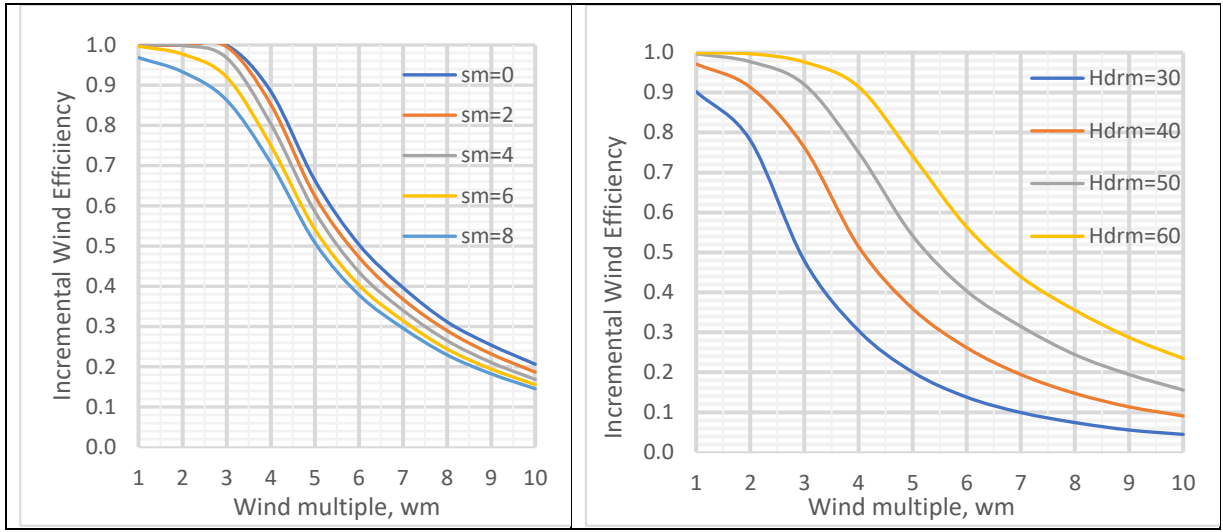

The equal spacing of the IWE vs wm curves in Figure 2 (left) for sm=0, 2, 4, 6 and 8, suggests that it should be unnecessary to model intermediate sm values in the IWE array. Minimum economically acceptable IWE and ISE values are likely to lie between and 0.3 and 0.7, and Figure 2 left shows IWE vs wm curves being significantly non- linear between IWE= 0.5 and 0.3. To minimise computational errors when interpolating between adjacent IWE values in the array, it will be necessary therefore to include IWE values in the array for all integer values between 1 and 10. Figure 2 right shows that in the IWE range of main interest, 0.3 to 0.7, the IWE vs wm curves for Hdrm values of 30, 40, 50 and 60 are remarkably equally spaced. This means that it should only be necessary to include in the IWE array Hdrm values of 30 , 40 and 50, and it may be possible to calculate IWE values for intermediate Hdrm values by interpolation, and up to Hdrm=60 by extrapolation.

Luckily it will be possible to test these hypotheses by checking with the Compound Model that the interpolated and extrapolated values result in the calculated IWE and ISE values being as anticipated. The Compound Model was used to calculate all IWE values for all combinations of Hdrm= 30, 40 and 50, wm=1 to 10 and sm= 0, 2, 4, 6 & 8, leading to the creation of the 150 element arrays of GW w+s and IWE values of Appendix 2. Bearing in mind that each IWE evaluation requires two Compound Model evaluations, 300 such evaluations were required to create these array. This illustrates the benefit of having the graphs of Figure 2 to guide the choice of Hdrm, wm and sm intervals. If without the insights gained from Figure 2 we had chosen to model say 5 Hdrm values, 10 wm values and 10 wm values, the creation of the IWE array would have required 1000 Compound Model predictions. In addition to the GW w+s and IWE arrays, Appendix 2 also includes what we have called an Investment Planning Table, which was derived from the IWE array.

The Investment Planning Table comprises 15 blocks of identical format for different combinations of Hdrm=30 ,40 and 50 and IWE = 0.7, 0.5, and 0.3. Since all 15 blocks are derived from the IWE array in the same way, it is sufficient to explain the creation of the entries in a single block, the Hdrm= 50 IWE=0.5 block being chosen for illustrative purposes. The first step was to identify in the IWE array the entries for Hdrm=50 which straddle IWE=0.5 for each sm value, these being highlighted in green. As an example for sm=8, IWE= 0.509 for wm=5, and IWE=0.379 for wm=6. By interpolation, IWE=0.5 for wm=5.068. This is what is entered in wm row of the Investment Planning Table for Hdrm 50 IWE=0.5 and sm=8. For each of the wm values calculated in this way the Compound Model is then used to calculate the Incremental Solar Efficiency value, ISE. For Hdrm=50, wm= 5.068 and sm=8 ISE= 0.349, and this is what is entered in the ISE row directly below the wm=5.068.

## 4.2 The Investment Lookup Table

The Investment Planning Table may be used to calculate wm and sm values for any IWE/ISE combination and generate an Investment Lookup Table. Since UK on- shore wind and solar generations currently have similar levelised costs[13], we shall illustrative the calculation of wm and sm values for Hdrm=50 and IWE= ISE = 0.5 and Hdrm=50. The first step in calculating the wm and sm values for this combination of Hdrm and IWE/ISE values is to find in the appropriate Investment Planning Table in Appendix 2 the ISE entries, and associated wm values, which straddle ISE=0.5. These are

| sm  | sm $_n$ (4)      | sm $_{n+1}$ (6)    |
|-----|------------------|--------------------|
| wm  | wm $_n$ (5.572)  | wm $_{n+1}$ (5.303)|
| ISE | ISE $_n$ (0.632) | ISE $_{n+1}$ (0.490)|

Using this nomenclature, the algorithm to calculate wm and sm values for the ISE $_{target}$ value of 0.5 is:

sm= sm $_{n+1}$ - ( sm $_{n+1}$ - sm $_n$) * ISE ratio
wm= wm $_{n+1}$ + Δ wm* ISE ratio
where
Δ ISE$_1$ = ISE$_n$ - ISE$_{n+1}$
Δ ISE$_2$ = ISE$_{target}$ - ISE$_{n+1}$
ISE ratio= Δ ISE$_2$ / Δ ISE$_1$
Δ wm= wm $_{n+1}$ -wm $_n$
 sm $_{n+1}$ - sm $_n$ is the sm sampling interval, which = 2.

The calculated results, wm= 5.322 and sm=5.859, and are entered in Investment Lookup Table 2, where they are highlighted in green. Repeating this process for all the entries in the Investment Planning Table results in the Hdrm=30, 40 and 50 entries in the Investment Lookup Tables 1, 2 and 3 which tabulate maximum wm and sm values for IWE/ISE =0.3, 0.5 and 0.7. Entries for Hdrm=60 are calculated by extrapolation and for Hdrm= 35, 45, 50 and 55 by interpolation. The bottom two rows are IWE and ISE values in each lookup table recalculated by the Compound Model using the Hdrm, wm and sm values above, and validate the approach taken.

**Table 1. Investment Lookup Table for IWE=ISE=0.3**

| Hdrm | 30    | 40    | 50    | 35    | 45    | 55    | 60    |
|------|-------|-------|-------|-------|-------|-------|-------|
| wm   | 4.257 | 5.661 | 7.085 | 4.959 | 6.673 | 7.798 | 8.510 |
| sm   | 4.207 | 5.585 | 6.977 | 4.959 | 6.281 | 7.672 | 8.368 |
| IWE  | 0.295 | 0.298 | 0.299 | 0.297 | 0.299 | 0.298 | 0.299 |
| ISE  | 0.296 | 0.300 | 0.300 | 0.298 | 0.300 | 0.295 | 0.301 |

**Table 2. Investment Lookup Table for IWE=ISE=0.5**

| Hdrm | 30    | 40    | 50    | 35    | 45    | 55    | 60    |
|------|-------|-------|-------|-------|-------|-------|-------|
| wm   | 3.209 | 4.263 | 5.322 | 3.736 | 4.792 | 5.851 | 6.380 |
| sm   | 3.457 | 4.705 | 5.859 | 4.081 | 4.792 | 6.437 | 7.014 |
| IWE  | 0.488 | 0.492 | 0.495 | 0.490 | 0.493 | 0.496 | 0.497 |
| ISE  | 0.504 | 0.497 | 0.500 | 0.500 | 0.499 | 0.501 | 0.503 |

**Table 3 Investment Lookup Table for IWE=ISE=0.7**

| Hdrm | 30    | 40    | 50    | 35    | 45    | 55    | 60    |
|------|-------|-------|-------|-------|-------|-------|-------|
| wm   | 2.611 | 3.487 | 4.387 | 3.049 | 3.937 | 4.838 | 5.288 |
| sm   | 2.775 | 3.926 | 4.762 | 3.351 | 4.344 | 5.181 | 5.599 |
| IWE  | 0.694 | 0.690 | 0.690 | 0.692 | 0.690 | 0.690 | 0.690 |
| ISE  | 0.725 | 0.702 | 0.711 | 0.711 | 0.707 | 0.715 | 0.719 |

The Investment Lookup Tables 1, 2 and 3 may be expressed in graphical form, as in Figure 3. The fact that all, including those derived by interpolation and extrapolation, lie on the same IWE/ISE lines, is a result of the equal spacing of the IWE vs wm curves in Figure 1, right, and has important consequences for the applicability of the tables. Although derived from IWE arrays for the specific Hdrm values of 30, 40 and 50, it becomes possible to use them across the whole range of Hdrm values from 30 to 60 by interpolation and extrapolation, without the need for further reference to the arrays from which they were derived.

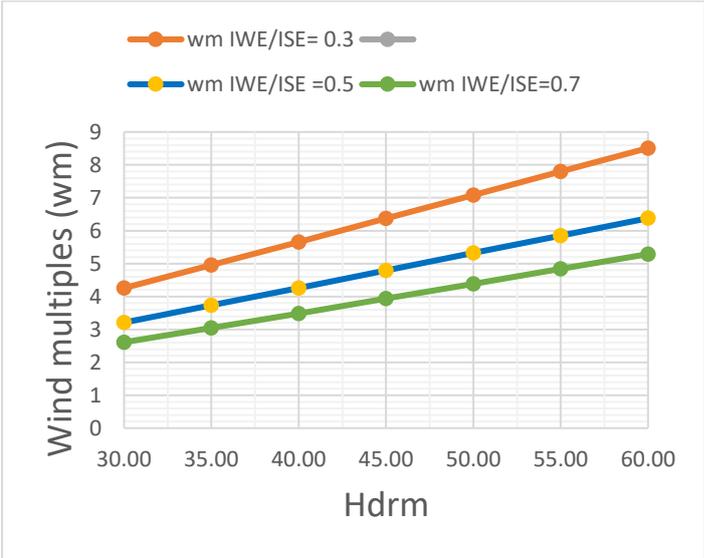

**Figure 3. wm predictions for Hdrm=30 to 60 and IWE=0.3, 0.5 and 0.7, derived from Tables 1, 2 and 3**

## 5. Application of the Investment Lookup Table

### 5.1 Predicting the appropriate level of wind capacity for 2030

An example of how the Lookup Tables might be used is in calculating the Hdrm needed in 2030 to justify the planned wind generation. On- shore wind capacity is currently around 15 GW and, with a load factor of 0.3, generates around 4.5 GW. Widespread resistance to the building of more on- shore turbines suggests that there will be little if any further investment by 2030. On the other hand, the government's target for offshore capacity in 2030 is 50GW by 2030. With a load factor of 0.45, this should generate around 22.5 GW, giving a total of around 27 GW, which is equivalent to wm=27/6.045=4.466. In the next section we shall consider the consequences of different IWE/ISE values, but in this example consider only IWE=ISE=0.5. Interpolation in Lookup Table 2 suggests a Hdrm of 41.92 would be needed to justify wm=4.466. We must consider the likelihood of the Hdrm being 41.92 in 2030. Electrical demand fell steadily by about 5 GW between 2013 and 2023 (Morley, 2023), and it is anticipated that future increases in electrical demand will mainly be in powering Electric Vehicles and electrically driven heat pumps for residential heating. However, sales of EVs and heat pumps are falling well below Climate Change Committee's targets (Wang, Wang and He, 2022). In 2023 demand on the electricity system was 29.6 GW and nuclear generation was 4.37 GW giving a Hdrm of 25.33 (Morley, 2023). It seems highly unlikely that Hdrm will increase to 41.92 by 2030, even bearing in mind an anticipated reduction in nuclear capacity. A further consideration is whether the necessary transmission infrastructure required to bring the additional offshore capacity to potential customers could be built by 2030. The National Grid has estimated that additional transmission infrastructure would cost of the order of £54Bn (Stacey and Harvey, 2024) but it is not clear how this will be financed. The National Grid has advised that its intention was to *"spend £19M improving infrastructure in 2025-*

*26"* (HM Government, 2021), and the Labour Party, which is expected to form the next UK government, advised in February 2024, that it anticipates spending £4.7Bn on green policies rather than the £28 Bn to which it had been previously committed. Nor is the availability of finance the only consideration. There has been very little investment in the UK's high voltage transmission system since the 1970's, and it is difficult to see where manpower with the appropriate skills might be found to carry out such a major project on such a short timescale.

### 5.2  Calculating the relationship between IWE/ISE values and residual carbon dioxide emissions

In addition to the Investment Lookup Tables being useful in analysing future electricity scenarios, they also provide a useful means of investigating the relationships between IWE/ISE values and carbon dioxide emissions. A Hdrm value of 48.5 GW was chosen for the five scenarios summarised in Table 4 in order to be able to make a direct comparison with the National Grid's 2035 scenario, for which Hdrm=48.5 (scenario E). Scenario A, which is also included for comparison purposes, shows carbon dioxide emissions if all the Hdrm of 48.5 GW were provided by combined cycle gas turbines which generate 4.87 Mtes pa carbon dioxide per GW deployed. The wm and sm values were calculated from the entries in Investment Lookup Table 2, by interpolation between Hdrm=40 and 50 for Hdrm=48.5 values. Wind plus solar generation entries, GW w+s, were calculated by interpolation in the GW w+s array in Appendix 1, and the Compound Model was used to recalculate the IWE and ISE values. These appear in the bottom two rows of Table 4, and give confidence in the calculation methodologies used.

**Table 4. Assessment of options for 2035 for Hdrm=48.5 and various IWE/ISE values**

| Scenario | A | B | C | D | E |
|---|---|---|---|---|---|
| IWE/ISE | n/a | 0.70 | 0.50 | 0.30 | 2035 scenario |
| wm | n/a | 4.25 | 5.16 | 6.87 | 8.96 |
| GWw (wm*6.045) | n/a | 25.71 | 31.21 | 41.54 | 54.16 |
| sm | n/a | 4.64 | 5.69 | 6.77 | 6.10 |
| GWs (sm*1.16) | n/a | 5.38 | 6.60 | 7.85 | 7.08 |
| Available wind+ solar | n/a | 31.08 | 37.80 | 49.39 | 61.24 |
| Accommodated wind + solar, GW w+s | n/a | 29.76 | 33.77 | 38.29 | 41.06 |
| Curtailed generation | n/a | 1.32 | 4.03 | 11.10 | 20.18 |
| Dispatchable generation | n/a | 18.74 | 14.73 | 10.21 | 7.44 |
| carbon dioxide emissions Mtes pa | 236.20 | 91.26 | 71.73 | 49.74 | 36.23 |
|  |  |  |  |  |  |
| IWE | n/a | 0.69 | 0.49 | 0.30 | 0.18 |
| ISE | n/a | 0.71 | 0.50 | 0.30 | 0.25 |

Table 5, which was derived from entries in table 4, shows the declining efficiency with which wind and solar generations are able to decarbonise the electricity system, as it proceeds from scenario A to scenario E in Table 4. The maximum theoretical decarbonisation efficiency is for small wind and solar fleets, for which all wind and solar generations are accommodated by the grid, with a reduction of 4.87 Mtes pa carbon dioxide per GW deployed. It is not possible to offer a view on what the minimum economically acceptable IWE and ISE values will be, since they will depend on the unknown unit costs of other means of reducing carbon dioxide emissions. It is reasonable however to suggest that IWE/ISE values of less than 0.5 would be unacceptable; at IWE/ISE=0.5 half of incremental increases in wind and solar generations would be curtailed, and unit costs of generation would be doubled. The National Grid's 2035 scenario, for which IWE= 0.18 and ISE= 0.25, suggests that, of addition increments of wind and solar generation, respectively 82 % and 75 % would be curtailed.

Table 5. Analysis of the declining efficiency with which the wind and solar fleets are able to decarbonise the electricity system as they move progressively from scenario A to E

| Between scenarios | Reduction in Mtes pa carbon dioxide emissions | Additional wind and solar generation | carbon dioxide emission reduction per GW wind and solar generation between scenarios |
|---|---|---|---|
| A & B | 144.94 | 31.08 | 4.66 |
| B & C | 19.53 | 6.72 | 2.91 |
| C & D | 21.99 | 11.59 | 1.9 |
| D & E | 13.5 | 11.85 | 1.14 |

## 6. Concluding Remarks

Steady state models of energy systems such as the National Grid's scenario for the year 2035 in Figure 1 are superficially attractive because they appear to provide a pathway to net zero by simply increasing wind and solar generations. Unfortunately, such models overlook the reality that the highly variable nature of wind and solar generations lead to an increasing amount of excess generation as the wind and solar fleets increase in size which cannot be put to beneficial use. Curtailed excess generation leads to a progressive reduction of the efficiency of the wind and solar fleets which eventually limit their deployment. Steady state analyses, by definition, are unable to address such issues. The authors have previously described the development of a Compound Model which incorporates appropriately scaled real time historic wind and solar generation records, providing an understanding of both the likely dynamic nature of future energy systems and, when averaged, the wind and solar fleet efficiencies which are likely to determine the upper limit of their deployment.

This paper takes this modelling approach further, showing how it is possible to determine from these efficiencies, IWE and ISE, the upper limit of the deployment of the wind and solar fleets for the range of electricity system scenarios likely to be encountered in the medium term future. Although we cannot know what the lowest economically acceptable IWE and ISE values will be, because they will depend to some extent on the costs of other means of reducing the country's greenhouse gas emissions, they are unlikely to be less than 0.5. If the IWE and ISE values are around 0.5, wind and solar generation will only be able to decarbonise around 70 % of the electricity system. The other 30% is likely to be provided by combined cycle gas turbines with sufficient capacity to provide all of the electricity demand during winter wind lulls. Combined cycle gas turbine generation is probably the only source of generation which is fast enough acting to mitigate the occasional rapid reductions in wind and solar generation studied by MacKay in the Republic of Ireland's wind generation records for 2007 (MacKay, 2009). Similar sized wind slews to those identified by MacKay are identified in the dynamic models of a 2035 scenario in Appendix 2.

The modelling approach described in this paper involves first using the Compound Model to create arrays over the range of operating conditions likely to be encountered by electricity system. Two stages of interpolation are described which lead to the creation of Investment Lookup Tables. These provide a means of calculating how much wind and solar generation will be justified economically, depending on the demand on the electricity system. The Investment Lookup Tables suggest that the electricity demand in 2030 might not be sufficient to justify the government's target for 2030 of 50 GW of offshore wind capacity.

**Appendix 1 Predicting the dynamic behaviour of the wind and solar fleets in 2035**

Although the main focus of the research described in this paper was on developing methods of predicting the upper economic sizes of future wind and solar fleets, understanding the dynamic interactions between wind and solar generations and the electricity system is also extremely important. Luckily, embedded in the Compound Model are weekly dynamic sub-models which give insights into the dynamic behaviour of the electricity system for the whole range of meteorological conditions the wind and solar fleets are likely to experience during a year. In this section we shall look at the model predictions for 2035, assuming Hdrm=48.5, wm=8.96 and sm=6.1. The models are based on appropriately scaled records of week 2, 24 and 52 2017, chosen because the predictions for 2035 illustrate different aspects of the interactions between the wind and solar fleets and the electricity system. Currently the demand on the grid varies by around +/- 10 GW each day, but it is assumed that by 2035 there will be a sufficient number of EVs with V2G capability to create a level demand [25]. In each of the figures Hdrm is shown as a blue line (at 48.5 GW), wind generation as green, solar generation as orange, and wind plus solar generation as a grey dotted line. When wind plus solar generation exceeds Hdrm, there is excess generation which must be curtailed and, when less than Hdrm, there is a deficit which has to be made good by other means.

**Figure 4 Simulation of electricity system in 2035, based on week 3 2017 records**

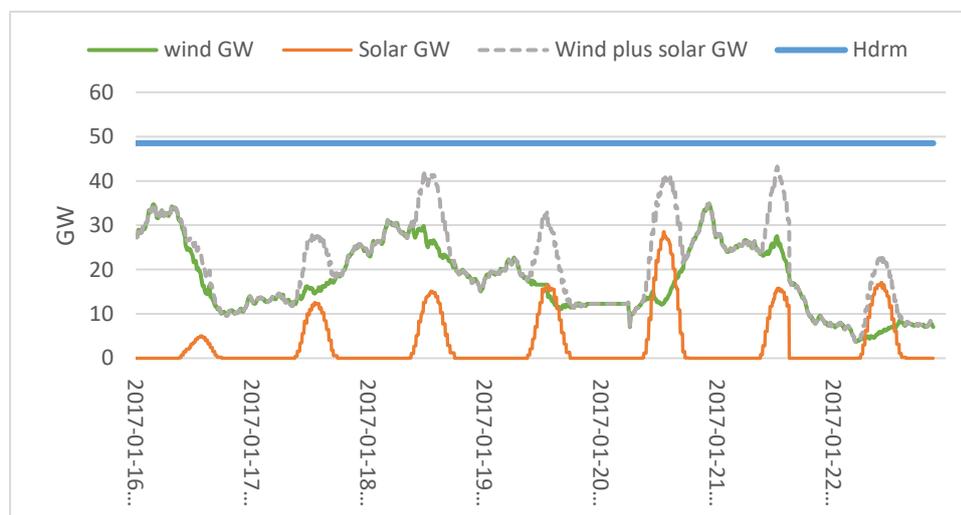

In January 2017 high pressure system was stationary over the UK and Europe, and as far as the Middle East, resulted in low temperatures and low winds. Although the UK is normally able to count on importing energy from Europe, particularly French surplus nuclear energy, the French and German records show that there was no surplus energy available to import[22]. The model suggests that if the 10 days wind lull, which spanned weeks 3 and 4 in 2017, were to reoccur in 2035 the average deficit would be 27.42 GW and the maximum deficit 44.9 GW. The UK has become increasingly reliant on imported energy in recent years, and the intercountry transfer records of 2017 suggest it would be foolhardy for the UK to rely on imports at times during winter wind lulls[22]. It is sometimes suggested that stored energy might be used to mitigate energy deficits caused by winter wind lulls, but the models do not support this suggestion. The energy deficit over the 10 day period in 2035 would be 6580 GWh, while the National Grid anticipates only 140 GWh of stored energy being available in 2035[3], including 27 GWh in the UK's four pumped storage reservoirs. MacKay [24] estimated that an additional 400 GWh might be stored if all potential Scottish valleys were flooded but by far the largest potential source of stored energy in 2035 is likely to be around 1050 GWh in Electric Vehicles with V2G capability. Their owners are however unlikely to be enthusiastic to restoring more than a small proportion of this energy if there was no prospect of it being returned within a reasonable time. As is currently the case, it is likely that combined cycle gas turbines generation will be needed to mitigate winter wind lull deficits. The maximum deficit in Figure 4 is 44.8 GW on 22nd January, suggests that it would be prudent to have a reserve fleet of combined cycle gas turbines with a capacity of around 50 GW. Assuming 0.5 is the lowest economically acceptable IWE/ISE value, table 4 suggests the need for an annual average dispatchable generation of 14.73 GW, which would result in an annual average utilisation of the ccgts of just under 30%.

**Figure 5 Simulation of electricity system in 2035, based on week 24 2017 records**

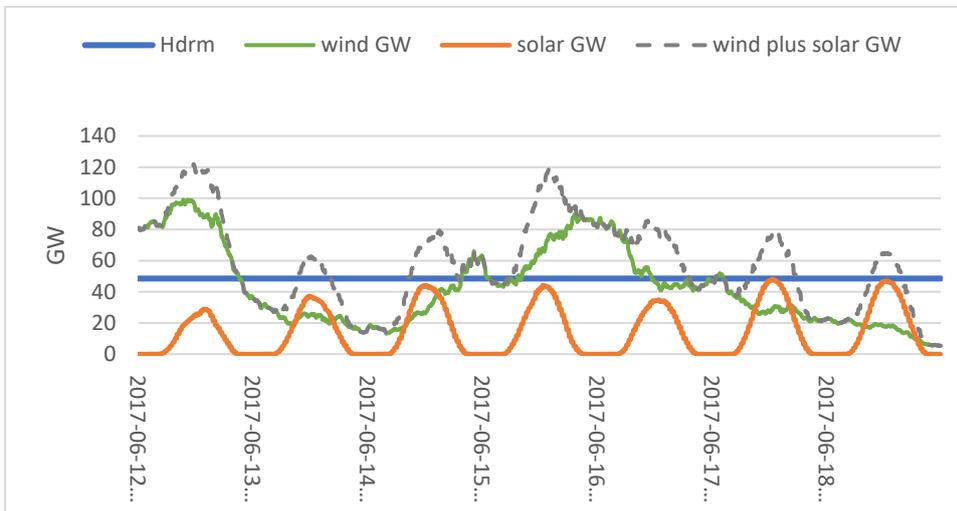

The simulation for based on week 24 2017 records provides a useful illustration of the consequences of wind and solar generations sharing the same Hdrm. For most of the week wind generation alone would not have exceeded Hdrm, but the combination of wind and solar generations results in the need for some curtailment of excess every day. Following his study of Republic of Ireland wind records for 11 Feb 2007, MacKay [24] suggested the UK might in future experience wind slews, rapid reductions in wind generation, of the order of 0.37 GW per GW of available wind generation, equivalent to a slew rate 22.63 GW/hr for the 2035 scenario. MacKay was only concerned about possible rapid reductions in wind generation but the problem is now exacerbated by having to consider coincident reductions in both wind and solar generation. Figure 5 shows such a combined wind plus solar slew on 12th June, falling from 86.1 GW at 17.40 to 52.3 GW at 2040 a slew rate of 11.26 GW/hr.

Figure 6 **Simulation of electricity system in 2035, based on week 52 2017 records**

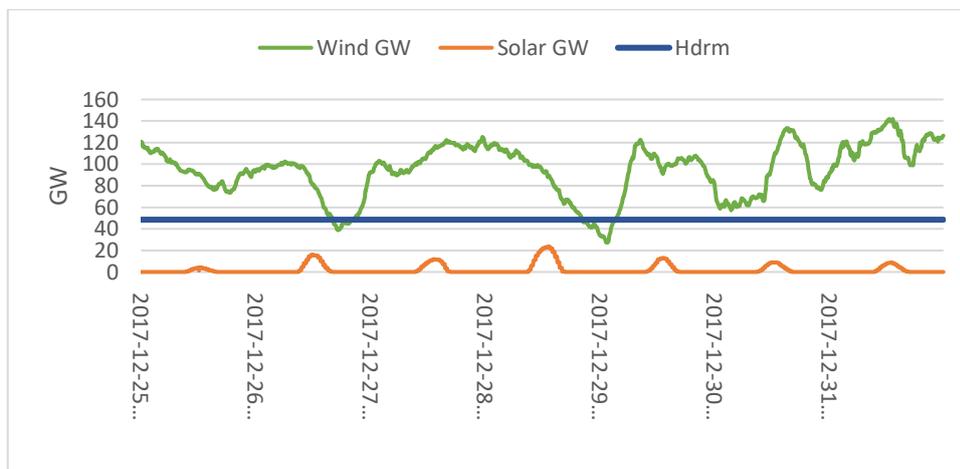

Christmas week in 2017, week 52, was a week of gales. If repeated in 2035, the average excess generation would be 48.34 GW, or an excess during the week of 8121 GWh. Excess generation peaked at 101.7 GW on New Year's Eve, and on the previous evening wind generation fell by 18.68 GW, between 7.20 and 8.05, a slew rate of 24.9 GW/h, not dissimilar to a 22.63 GW/h slew rate we might expect, based on MacKay's observation of a wind slew in the 2007 in the Republic of Ireland.

The predictions for the three weeks just analysed in 2035, based on 2017 records, highlight the importance of analysing future electricity systems dynamically. During a 10-day winter wind lull such as that experienced in weeks 3 and 4 of 2017 we would expect an energy deficit over the 10 day period of around 6580 GWh. The peak deficit of 44.7 GW suggests that, realistically, it will be necessary to mitigate such as wind lull with a fleet of ccgts capable of generating around 50 GW. The ccgts would have an annual utilisation of only around 15%, and would therefore have to be regarded, and paid for, as strategic insurance. The slew rates of 11.26 GW/h in Figure 4 and 22.63 GW/h in Figure 5 suggest particular attention will have to be paid to electricity system stability as the wind fleets increase in size.

# Appendix 2 GW w+s and IWE Arrays and INVESTMENT PLANNING TABLE

## GW w+s ARRAY ( Wind and solar generation accomodated by the electricity system)

### Hdrm=30

| wm | sm=0 | sm=2 | sm=4 | sm=6 | sm=8 |
|----|------|------|------|------|------|
| 1  | 6.05 | 8.364 | 10.6 | 12.27 | 13.3 |
| 2  | 12.1 | 14.36 | 16.3 | 17.44 | 18.1 |
| 3  | 17.2 | 19.13 | 20.6 | 21.35 | 21.8 |
| 4  | 20.5 | 22.05 | 23.1 | 23.71 | 24 |
| 5  | 22.7 | 23.95 | 24.8 | 25.24 | 25.5 |
| 6  | 24.2 | 25.24 | 25.9 | 26.28 | 26.5 |
| 7  | 25.2 | 26.14 | 26.7 | 26.99 | 27.2 |
| 8  | 26   | 26.8  | 27.3 | 27.52 | 27.7 |
| 9  | 26.6 | 27.3  | 27.7 | 27.92 | 28 |
| 10 | 27.1 | 27.69 | 28   | 28.22 | 28.3 |

### Hdrm=40

| wm | sm=0 | sm=2 | sm=4 | sm=6 | sm=8 |
|----|------|------|------|------|------|
| 1  | 6.05 | 8.364 | 10.7 | 12.9 | 14.5 |
| 2  | 12.1 | 14.41 | 16.7 | 18.6 | 19.9 |
| 3  | 18.1 | 20.31 | 22.3 | 23.8 | 24.8 |
| 4  | 23   | 24.91 | 26.6 | 27.7 | 28.5 |
| 5  | 26.4 | 28.09 | 29.4 | 30.4 | 30.9 |
| 6  | 28.9 | 30.38 | 31.5 | 32.3 | 32.7 |
| 7  | 30.8 | 32.07 | 33   | 33.7 | 34.1 |
| 8  | 32.2 | 33.35 | 34.2 | 34.7 | 35 |
| 9  | 33.3 | 34.32 | 35   | 35.5 | 35.8 |
| 10 | 34.2 | 35.09 | 35.7 | 36.1 | 36.4 |

### Hdrm=50

| wm | sm=0 | sm=2 | sm=4 | sm=6 | sm=8 |
|----|------|------|------|------|------|
| 1  | 6.045 | 8.364 | 10.7 | 13   | 15.1 |
| 2  | 12.09 | 14.41 | 16.7 | 19   | 20.9 |
| 3  | 18.14 | 20.45 | 22.7 | 24.8 | 26.3 |
| 4  | 23.99 | 26.18 | 28.2 | 29.9 | 31.2 |
| 5  | 28.7  | 30.68 | 32.4 | 33.9 | 34.9 |
| 6  | 32.25 | 34.02 | 35.5 | 36.7 | 37.6 |
| 7  | 35    | 36.57 | 37.9 | 38.9 | 39.6 |
| 8  | 37.15 | 38.57 | 39.7 | 40.6 | 41.2 |
| 9  | 38.87 | 40.16 | 41.2 | 42   | 42.5 |
| 10 | 40.28 | 41.44 | 42.4 | 43   | 43.5 |

## INCREMENTAL WIND EFFICIENCY ARRAY (IWE)

### Hdrm=30

| wm | sm=0 | sm=2 | sm=4 | sm=6 | sm=8 |
|----|------|------|------|------|------|
| 1  | 0.999 | 0.999 | 0.975 | 0.902 | 0.840 |
| 2  | 0.985 | 0.947 | 0.855 | 0.781 | 0.734 |
| 3  | 0.658 | 0.595 | 0.524 | 0.479 | 0.455 |
| 4  | 0.425 | 0.375 | 0.332 | 0.304 | 0.289 |
| 5  | 0.288 | 0.250 | 0.219 | 0.201 | 0.190 |
| 6  | 0.205 | 0.173 | 0.151 | 0.138 | 0.131 |
| 7  | 0.149 | 0.124 | 0.108 | 0.099 | 0.094 |
| 8  | 0.111 | 0.093 | 0.081 | 0.074 | 0.071 |
| 9  | 0.084 | 0.070 | 0.061 | 0.056 | 0.053 |
| 10 | 0.067 | 0.057 | 0.049 | 0.045 | 0.042 |

### Hdrm=40

| wm | sm=0 | sm=2 | sm=4 | sm=6 | sm=8 |
|----|------|------|------|------|------|
| 1  | 0.999 | 0.999 | 0.999 | 0.971 | 0.923 |
| 2  | 0.999 | 0.999 | 0.975 | 0.913 | 0.854 |
| 3  | 0.929 | 0.891 | 0.829 | 0.762 | 0.716 |
| 4  | 0.661 | 0.614 | 0.560 | 0.515 | 0.482 |
| 5  | 0.474 | 0.432 | 0.391 | 0.359 | 0.339 |
| 6  | 0.349 | 0.318 | 0.286 | 0.262 | 0.247 |
| 7  | 0.266 | 0.238 | 0.213 | 0.194 | 0.184 |
| 8  | 0.206 | 0.181 | 0.161 | 0.147 | 0.138 |
| 9  | 0.161 | 0.140 | 0.125 | 0.114 | 0.107 |
| 10 | 0.128 | 0.111 | 0.100 | 0.091 | 0.086 |

### Hdrm=50

| wm | sm=0 | sm=2 | sm=4 | sm=6 | sm=8 |
|----|------|------|------|------|------|
| 1  | 0.999 | 0.999 | 0.999 | 0.997 | 0.968 |
| 2  | 0.999 | 0.999 | 0.999 | 0.977 | 0.933 |
| 3  | 0.999 | 0.996 | 0.968 | 0.920 | 0.863 |
| 4  | 0.885 | 0.853 | 0.804 | 0.751 | 0.707 |
| 5  | 0.664 | 0.625 | 0.586 | 0.542 | 0.509 |
| 6  | 0.504 | 0.473 | 0.436 | 0.403 | 0.379 |
| 7  | 0.397 | 0.368 | 0.340 | 0.315 | 0.296 |
| 8  | 0.312 | 0.290 | 0.265 | 0.244 | 0.230 |
| 9  | 0.254 | 0.232 | 0.211 | 0.195 | 0.183 |
| 10 | 0.206 | 0.187 | 0.169 | 0.156 | 0.145 |

## INVESTMENT PLANNING TABLE

### Hdrm=30  IWE=0.7

|     | sm=0 | sm=2 | sm=4 | sm=6 | sm=8 |
|-----|------|------|------|------|------|
| wm  | 2.871 | 2.701 | 2.469 | 2.267 | 2.120 |
| ISE | 0.902 | 0.796 | 0.548 | 0.326 | 0.215 |

### Hdrm=40  IWE=0.7

|     | sm=0 | sm=2 | sm=4 | sm=6 | sm=8 |
|-----|------|------|------|------|------|
| wm  | 3.856 | 3.690 | 3.479 | 3.252 | 3.070 |
| ISE | 0.900 | 0.827 | 0.695 | 0.477 | 0.372 |

### Hdrm=50  IWE=0.7

|     | sm=0 | sm=2 | sm=4 | sm=6 | sm=8 |
|-----|------|------|------|------|------|
| wm  | 4.835 | 4.670 | 4.477 | 4.24 | 4.03 |
| ISE | 0.899 | 0.839 | 0.762 | 0.6 | 0.440 |

### Hdrm: IWE=0.5

|     | sm=0 | sm=2 | sm=4 | sm=6 | sm=8 |
|-----|------|------|------|------|------|
| wm  | 3.678 | 3.432 | 3.126 | 2.930 | 2.838 |
| ISE | 0.781 | 0.664 | 0.439 | 0.248 | 0.164 |

### Hdrm: IWE=0.5

|     | sm=0 | sm=2 | sm=4 | sm=6 | sm=8 |
|-----|------|------|------|------|------|
| wm  | 4.863 | 4.626 | 4.354 | 4.095 | 3.923 |
| ISE | 0.784 | 0.708 | 0.562 | 0.385 | 0.25 |

### Hdrm=50  IWE=0.5

|     | sm=0 | sm=2 | sm=4 | sm=6 | sm=8 |
|-----|------|------|------|------|------|
| wm  | 6.026 | 5.820 | 5.572 | 5.303 | 5.07 |
| ISE | 0.788 | 0.735 | 0.632 | 0.490 | 0.35 |

### Hdrm: IWE=0.3

|     | sm=0 | sm=2 | sm=4 | sm=6 | sm=8 |
|-----|------|------|------|------|------|
| wm  | 4.914 | 4.602 | 4.282 | 4.043 | 3.884 |
| ISE | 0.64 | 0.503 | 0.31 | 0.176 | 0.12 |

### Hdrm: IWE=0.3

|     | sm0 | sm2 | sm4 | sm6 | sm8 |
|-----|-----|-----|-----|-----|-----|
| wm  | 6.591 | 6.225 | 5.865 | 5.607 | 5.42 |
| ISE | 0.63 | 0.549 | 0.42 | 0.27 | 0.18 |

### Hdrm: IWE=0.3

|     | sm=0 | sm=2 | sm=4 | sm=6 | sm=8 |
|-----|------|------|------|------|------|
| wm  | 8.204 | 7.870 | 7.537 | 7.21 | 6.95 |
| ISE | 0.635 | 0.571 | 0.47 | 0.35 | 0.24 |